\documentclass[a4paper]{jpconf}
\usepackage{psfig}
\usepackage{amsmath}
\usepackage{amssymb}
\usepackage{graphicx}

\newcommand{\lSect}[1]{{\label{sec:#1}}}

\newcommand{\Sectff}[1]{{\ref{sec:#1}}}
\newcommand{\Sect}[1]{{\S~\Sectff{#1}}}

\def\gtaprx {\lower .1ex\hbox{\rlap{\raise .6ex\hbox{\hskip .3ex
    {\ifmmode{\scriptscriptstyle >}\else
        {$\scriptscriptstyle >$}\fi}}}
    \kern -.4ex{\ifmmode{\scriptscriptstyle \sim}\else
        {$\scriptscriptstyle\sim$}\fi}}}
\def\ltaprx {\lower .1ex\hbox{\rlap{\raise .6ex\hbox{\hskip .3ex
    {\ifmmode{\scriptscriptstyle <}\else
        {$\scriptscriptstyle <$}\fi}}}
    \kern -.4ex{\ifmmode{\scriptscriptstyle \sim}\else
        {$\scriptscriptstyle\sim$}\fi}}}

\newcommand{\note}[1]{\emph{\textcolor{red}{}}}

\newcommand{\Msun}{{\ensuremath{\mathrm{M}_{\odot}}}}

\newcommand{\CASTRO}{\texttt{CASTRO}}
\newcommand{\FLASH}{\texttt{FLASH}}
\newcommand{\KEPLER}{\texttt{KEPLER}}
\newcommand{\MESA}{\texttt{MESA}}

\makeatletter

\newcommand{\Rmnum}[1]{\expandafter\@slowromancap\romannumeral #1@}
\makeatother

\begin{document}
\title{Conservative Initial Mapping For Multidimensional Simulations of Stellar Explosions}
\author{Ke-Jung Chen$^1$, Alexander Heger$^1$, and Ann Almgren$^{2}$}

\address{$^1$ Minnesota Institute for Astrophysics, University of Minnesota, Minneapolis, MN 55455}
\address{$^2$ Center for Computational Sciences and Engineering, Lawrence Berkeley National Lab, Berkeley, CA 94720}

\ead{kchen@physics.umn.edu}

\begin{abstract}
Mapping one-dimensional stellar profiles onto multidimensional grids as initial conditions for hydrodynamics 
calculations can lead to numerical artifacts, one of the most severe of which is the violation of conservation 
laws for physical quantities such as energy and mass.  Here we introduce a  
numerical scheme for mapping one-dimensional spherically-symmetric data onto multidimensional meshes
so that these physical quantities are conserved. We validate our scheme by porting a realistic 1D Lagrangian
stellar profile to the new multidimensional Eulerian hydro code \CASTRO.  Our results show that all important 
features in the profiles are reproduced on the new grid and that conservation laws are enforced at all 
resolutions after mapping.
\end{abstract}

\section{Introduction}
Multidimensional simulations shed light on how fluid instabilities arising in supernova explosions mix ejecta 
\cite{herant1994,candace2009,candace2010,candace2011}. Unfortunately, computing the full self-consistent 
three-dimensional (3D) stellar evolution initial models for the explosion setup is still beyond the 
realm of contemporary computational power. One alternative is to first evolve the main sequence star in a 1D 
stellar evolution code in which the equations of momentum, energy and mass are solved on a spherically 
symmetric Lagrangian grid, such as \KEPLER{} \cite{kepler} or \MESA{} \cite{mesa}. Once the star reaches the 
pre-supernova phase, its 1D profiles can then be mapped into multidimensional hydro codes such as \CASTRO{} 
\cite{zhang2011,ann2010} or \FLASH{} \cite{flash} and continue to be evolved until the star explodes.

Differences between codes in dimensionality and coordinate mesh can lead to numerical issues such 
as violation of conservation of mass and energy when profiles are mapped from one code to another.  A first, simple
approach could be to initialize multidimensional grids by linear interpolation from corresponding mesh points 
on the 1D profiles. However, linear interpolation becomes invalid when the new grid fails to resolve critical 
features in the original profile such as the inner core of a star.  This is especially true when porting profiles 
from 1D Lagrangian codes, which can easily resolve very small spatial features in mass coordinate, to a 
fixed or adaptive Eulerian grid.  Besides conservation laws, some physical processes such as nuclear 
burning are very sensitive to temperature, so slight errors in mapping can lead to very different outcomes 
for the simulation. Only a few studies have examined mapping 1D profiles to 2D or 3D meshes 
\cite{zingale2002}, and none address the conservation of physical quantities by such procedures. 

We investigate these issues and introduce a new scheme for mapping 1D data sets to multidimensional 
grids.  We first describe our mapping algorithm in \Sect{method} and then present results of porting a 
massive star model from \KEPLER{} to \CASTRO{} in \Sect{result}. Finally, we conclude in \Sect{conclusion}. 

\section{Method}
\lSect{method}
Since the star is very nearly in hydrostatic equilibrium but we also map explosions where we want to conserve 
the total energy , care must be taken in mapping its profile from the 
non-uniform Lagrangian grid in mass coordinate to the new Eulerian spatial grid. Our method preserves 
conservation of quantities such as mass and energy that are analytically conserved in the evolution 
equations on the new mesh.  Although this does not guarantee that a hydrostatic star will be fully hydrostatic 
on the new grid because our construction does not, it is a physically motivated constraint and sufficient for our simulations. The algorithm we describe 
below is specific to our stellar models but can be easily generalized to other mappings of 1D data to higher 
dimensions.

\begin{figure}
\begin{center} 
 \includegraphics[scale=0.6]{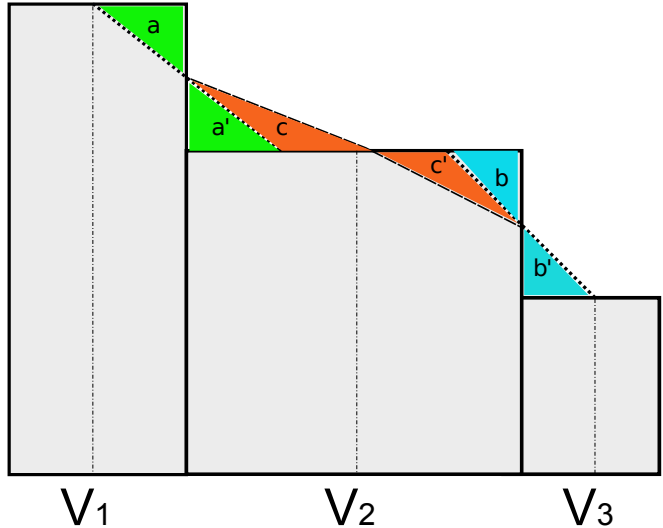}
  \caption{ Constructing a conservative profile. The rectangular bins illustrate the original 1D profile. The 
areas of different colors represent conserved quantities such as mass and internal energy. Note that 
uniform zones in mass lead to nonuniform bins in volume coordinate, as shown above.   \label{mapping}}
\end{center}
\end{figure}

\begin{figure}
\begin{center} 
 \includegraphics[scale=0.6]{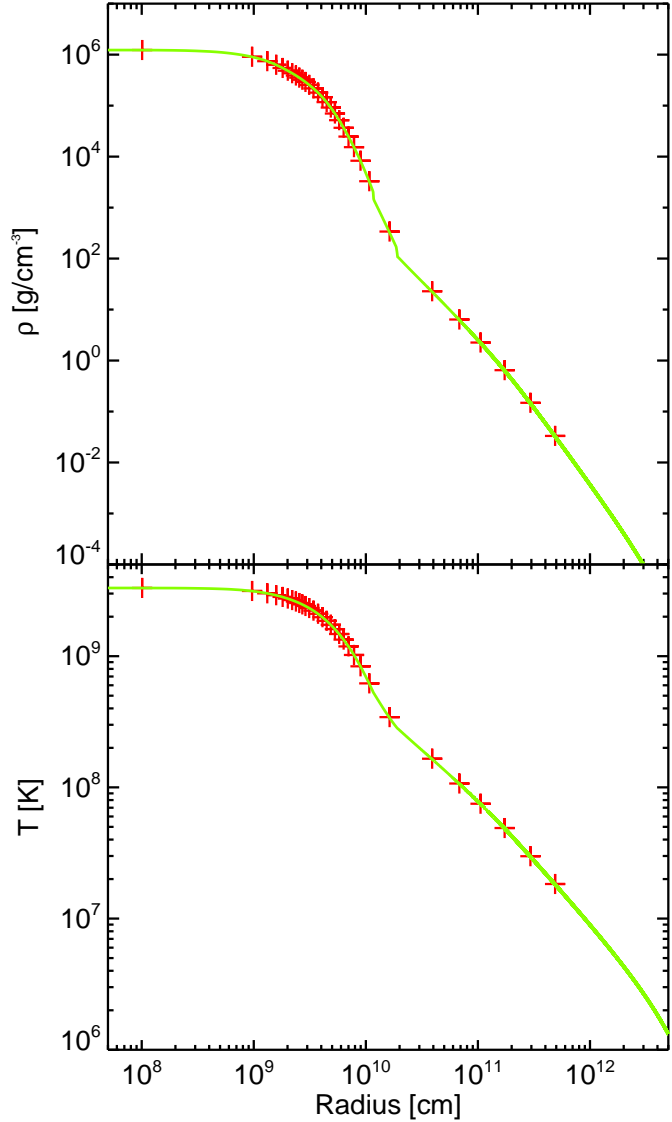}
  \caption{\KEPLER{} densities and temperatures (red crosses) and our piecewise linear fits (green lines). 
Since we map internal energy (a conserved quantity) rather than temperature, we calculate $T$ from the 
equation of state using density, element abundance, and internal energy. \label{fig1}}
\end{center}
\end{figure}

\begin{figure}
\begin{center} 
 \includegraphics[scale=0.6]{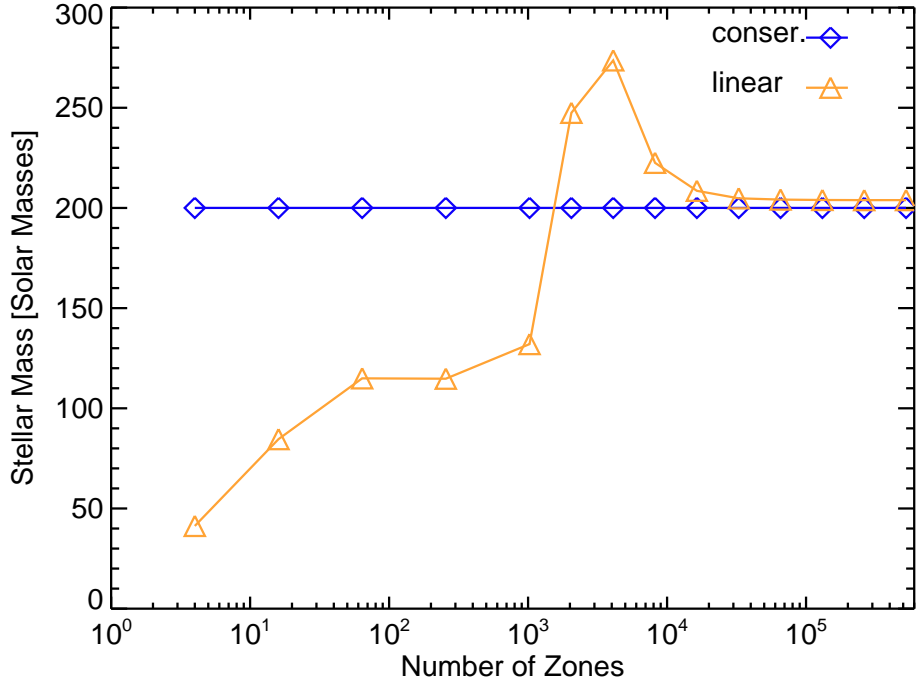}
  \caption{Total mass of the star on the new 1D \CASTRO{} grid vs. number of zones.  Conservative 
mapping (blue) preserves the mass of the star at all resolutions, while linear interpolation (orange)  
converges to 200 \Msun{} at a resolution of $\sim$ a few $\times 10^4$, when the grid begins to 
resolve the core of the star ($\sim 10^9$ cm). Even using a very high resolution, the linear interpolation 
still fails to resolve the density gradient between the inner core and the out envelope (see Figure \ref{fig1} 
$10^{9} \sim 10^{11}$ cm). So the curve of linear interpolation is saturated at zone number $\sim 10^5$, 
its results are still off by a few $\%$.  \label{fig2}}
\end{center}
\end{figure}

\begin{figure}
\begin{center} 
 \includegraphics[scale=0.6]{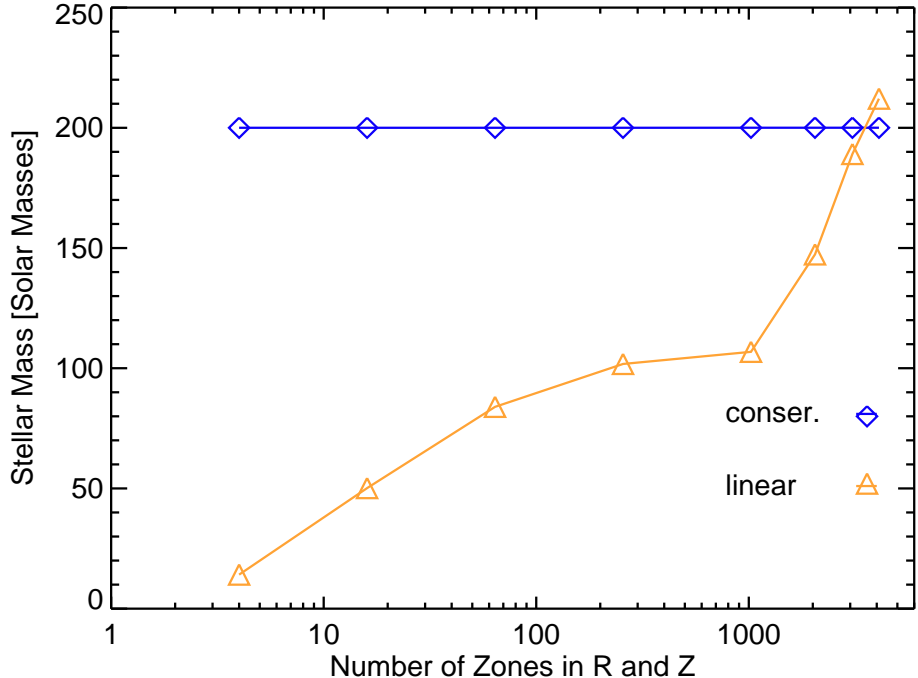}
  \caption{Total mass of the star on the new 2D \CASTRO{} grid vs. number of zones in both $r$ and 
$z$.  Conservative mapping (blue) recovers the mass of the star at all resolutions and linear 
interpolation (orange) approaches 200 \Msun{} at a resolution of $\sim$ $2048^2$. \label{fig3}}
\end{center}
\end{figure}

\begin{figure}
\begin{center} 
 \includegraphics[scale=0.6]{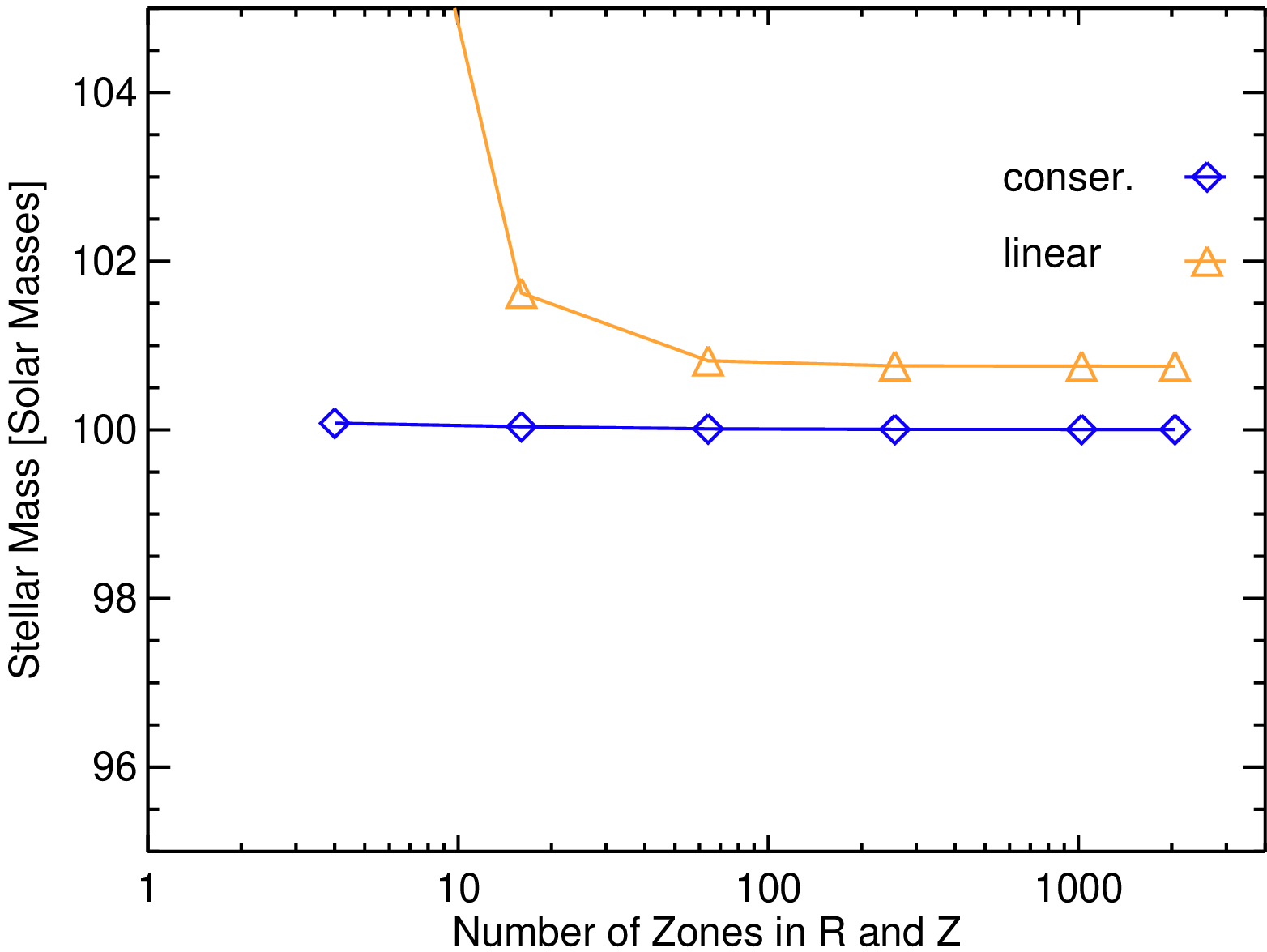}
  \caption{Total mass of the He core on the 2D \CASTRO{} grid vs. number of zones in both $r$ and 
$z$.  Conservative mapping (blue) preserves the original mass of the core at all resolutions while 
linear interpolation (orange) begins to converge to 100 \Msun{} at a resolution of $64^2$ but is still 
off by $\sim 1 \%$ even as the resolution approaches $\sim 2048^2$ .  \label{fig4}}
\end{center}
\end{figure}

First, we construct a continuous (C$^0$) function that conserves the physical quantity when it is mapped onto the 
new grid. An ideal choice for interpolation is the volume coordinate $V$, the volume enclosed by a given radius from the center of 
the star.  Then, integrating a density $\rho_X$ (which can represent mass or internal energy density) with respect to the volume coordinate yields a conserved quantity $X$
\begin{equation}
X=\int _{V_1}^{V_2}\rho_X\,\mathrm{d}V,
\end{equation}
such as the total mass or internal total energy lying in the shell between $V_1$ and $V_2$. 

Next, we define a piecewise linear function in volume $V$ that represents the conserved quantity $\rho_X$, 
preserves its monotonicity (no new artificial extrema), and is bounded by the extrema of the original data.  
The segments are constructed in two stages.  First, we extend a line across the interface between adjacent 
zones that either ends or begins at the center of the smaller of the two zones, as shown in Figure 
\ref{mapping} (note that uniform zones in mass coordinate do not result in uniform zones in $V$).  The 
slope of the segment is chosen so that the area trimmed from one zone by the segment ($a$ and $b$) is 
equal to the area added under the segment in the neighboring bin ($a'=a$ and $b'=b$).

If the two segments bounding $a$ and $a'$ and $b$ and $b'$ are joined together by a third in the center zone in 
Figure \ref{mapping}, two ``kinks'', or changes in slope, can arise in the interpolated quantity there; plus, the slope 
of the flat central segment is usually a poor approximation to the average gradient in that interval.  We therefore 
construct two new segments that span the entire central zone and connect with the two original segments where 
they cross its interfaces, as shown in Figure \ref{mapping}.  The new segments join each other at the position in 
the central bin where the areas $c$ and $c'$ enclosed by the two segments are equal (note that they in general 
have different slopes).  After repeating this procedure everywhere on the grid, each bin will be spanned by two 
linear segments that represent the interpolated quantity $\rho_X$ at any $V$ within the bin and have no more 
than one kink in $\rho_X$ across the zone.  Our scheme introduces some smearing (or smoothing) of the data, 
but it is limited to less than or equal to the width of one zone on the original grid.

The result of our interpolation scheme is a piecewise linear reconstruction in $V$ of the original profile in mass 
coordinate, for which the quantity $\rho_X$ can be determined at any $V$, not just the radii associated with the 
Lagrangian grid.  We show this profile as a function of the radius associated with the volume coordinate $V$ for 
a zero-metallicity 200 \Msun{} star with $r$ $\sim$ $2 \times 10^{13}$ cm from \KEPLER{} \cite{heger2002,heger2010} in Figure \ref{fig1}.

We populate the new multidimensional grid with conserved quantities from the reconstructed stellar profiles
as follows.  First, the distance of the selected mesh point from the center of the new coordinate grid is calculated.  
We then use this radius to obtain its $V$ to reference the corresponding density in the piecewise linear profile of 
the star. The density assigned to the zone is then determined from adaptive iterative subsampling.  This is done by first 
computing the total mass of the zone by multiplying its volume by the interpolated density. We then divide the 
zone into equal subvolumes whose sides are half the length of the original zone.  New $V$ are computed for 
the radii to the center of each of these subvolumes and their densities are again read in from the reconstructed 
profile.  The mass of each subvolume is then calculated by multiplying its interpolated density by its volume element.  These masses are then summed and 
compared to the mass previously calculated for the entire cell.  If the relative error between the two masses is larger 
than some predetermined tolerance, each subvolume is again divided as before, masses are computed for all the 
constituents comprising the original zone, and they are then summed and compared to the zone mass from the 
previous iteration.  This process continues recursively until the relative error in mass between the two most 
recent consecutive iterations falls within an acceptable value, typically 10$^{-4}$.  The density we assign to the 
zone is just this converged mass divided by the volume of the entire cell.  This method is used to map internal 
energy density and the partial densities of the chemical species to every zone on the new grid.  The total density 
is then obtained from the sum of the partial densities; pressure and temperature in turn are determined from the 
equation of state. This method is easily applied to hierarchy geometry of the target grid. 


\section{Results}
\lSect{result}
We port a 1D stellar model from  \KEPLER{} into \CASTRO{} to verify that our mapping is conservative. As an example here, we use the zero-metalicity pre-supernova 200 \Msun{} star whose profiles are shown in Figure \ref{fig1}.  \KEPLER{} is a 
Lagrangian code that evolves stars in mass coordinate so its mesh is nonuniform in space, and  \CASTRO{} has an 
Eulerian grid with uniform spatial zones. 

We compare our piecewise linear fits to the \KEPLER{} data in Figure \ref{fig1}, which shows that they reproduce 
the original stellar profile. Because our fits smoothly interpolate the block histogram structure of the \KEPLER{} bins 
(especially at larger radii), they reduce the number of unphysical sound waves that would have been introduced in 
\CASTRO{} by the discontinuous interfaces between these bins in the original data.  The density profile is key to the 
hydrodynamic and gravitational evolution of the explosion, and the temperature profile is crucial to the nuclear 
burning that powers the explosion.

We first map the profile onto a 1D grid in \CASTRO{} and plot the mass of the star as a function of grid resolution 
in Figure \ref{fig2}. The mass is independent of resolution for our conservative mapping because we subsample 
the quantity in each cell 
prior to initializing it, as described above. In contrast, the total mass from linear interpolation is very sensitive to 
the number of grid points but does eventually converge when the number of zones is sufficient to resolve the core 
of the star, in which most of its mass resides. 

We next map the \KEPLER{} profile onto a 2D cylindrical grid ($r,z$) and a 3D cartesian grid ($x,y,z$) in \CASTRO{}. 
The only difference between mapping to 1D, 2D, and 3D is the form of the volume elements used to subsample each 
cell,  which are $4\pi r^2 dr$, $2\pi r  dr dz$, $dx dy dz$, respectively.  We show the mass of the star as a 
function of resolution in Figure \ref{fig3}. Conservative mapping again preserves its mass at all grid 
resolutions. In 2D, more zones are required for linear interpolation to converge to the mass of the star.  To further 
validate our conservative scheme, we map just the helium core of star ($\sim$ 100 $\Msun$ with $r \sim$ 10$^{10}$ 
cm) onto the 2D grid. The helium core is crucial to modeling thermonuclear supernovae because it is where explosive 
burning begins.  We show its mass as a function of resolution in Figure \ref{fig4}.  We again recover 
all the mass of the core at all resolutions while linear interpolation overestimates the mass by at least $\sim$ 1 $\%$, 
even with large numbers of zones. 

Conservative mapping is effective in 3D but requires much more computational time to subsample each cell 
to convergence.  Furthermore, an impractical number of zones is needed for linear interpolation to reproduce the 
original mass of the star, so we defer its comparison to our scheme in 3D to a later study.  We note that our method 
also works with adaptive mesh refinement (AMR) grids because both $V$ and the interpolated quantities can be determined 
and subsampling can be performed on every grid in the hierarchy.
             
\section{Conclusion}
\lSect{conclusion}
Multidimensional stellar evolution and supernova simulations are numerically challenging because multiple physical 
processes (hydrodynamics, gravity, burning) occur on many scales in space and time.  For computational efficiency, 
1D stellar models are often used as initial conditions in 2D and 3D calculations.  Mapping 1D profiles onto 
multidimensional grids can introduce serious numerical artifacts, one of the most severe of which is the violation of 
conservation of physical quantities. We have developed a new mapping algorithm that guarantees that conserved 
quantities are preserved at any resolution and reproduces the most important features in the original profiles.  Our 
method is practical for 1D and 2D calculations, and we are now developing integral methods (an explicit integral approach instead of using volume subsampling) that are numerically tractable in 3D.   

\section*{Acknowledgments}
The authors thank Daniel Whalen for reviewing the earlier manuscript and providing many insightful comments, 
the members of the CCSE at LBNL for help with \CASTRO{}, and Hank Childs for assistance with {\it VisIt}.  We 
also thank Laurens Keek, Volker Bromm, Dan Kasen, Adam Burrows, and Stan Woosley for many useful discussions.  All 
numerical simulations were performed with allocations from the University of Minnesota Supercomputing Institute 
and the National Energy Research Scientific Computing Center. This work has been supported by the DOE SciDAC 
program under grants DOE-FC02-01ER41176, DOE-FC02-06ER41438, and DE-FC02-09ER41618, and by the US 
Department of Energy under grant DE-FG02-87ER40328.

\section*{References}
\bibliographystyle{iopart-num}
\bibliography{iopart-num}


\end{document}